\documentclass[MSNbibl,nameyear,dvips]{arxstspdf}
\usepackage{flushend}
\usepackage{stfloats}
\usepackage{graphicx}

\volume{27}
\issue{2}
\pubyear{2012}
\firstpage{199}
\lastpage{201}
\doi{10.1214/12-STS376REJ} 
\referstodoi{10.1214/11-STS376}

\makeatletter
\DeclareMathSymbol{\Real}{\mathbin}{AMSb}{"52}
\DeclareMathSymbol{\Natural}{\mathbin}{AMSb}{"4E}

\makeatother

\begin{document}
\begin{frontmatter}
\vspace*{6pt}
\title{Rejoinder}
\runtitle{Rejoinder}

\begin{aug}
\author[a]{\fnms{A.~C.} \snm{Davison}\corref{}\ead[label=e1]{Anthony.Davison@epfl.ch}},
\author[b]{\fnms{S.~A.} \snm{Padoan}\ead[label=e2]{Simone.Padoan@stat.unipd.it}}
\and
\author[c]{\fnms{M.} \snm{Ribatet}\ead[label=e3]{mathieu.ribatet@math.univ-montp2.fr}}
\runauthor{A.~C.~Davison, S.~A.~Padoan and M.~Ribatet}

\address[a]{Anthony
Davison is Professor, Chair of Statistics, Institute of
Mathematics, EPFL-FSB-IMA-STAT, Station 8, Ecole Polytechnique
F\'ed\'erale de Lausanne, 1015 Lausanne, Switzerland
\printead{e1}.}
\address[b]{Simone Padoan is a
Senior Assistant Researcher, Department of Statistical Science,
University of Padua, Via Cesare Battisti~241, 35121
Padova, Italy \printead{e2}.}
\address[c]{Mathieu Ribatet is a Ma\^itre
de conference, I3M, UMR CNRS 5149, Universite Montpellier II, 4
place Eugene Bataillon, 34095 Montpellier, cedex 5, France
\printead{e3}.}

\end{aug}



\end{frontmatter}

\section{Introduction}

We are grateful to the discussants for their posi\-tive and interesting
comments. In an area moving~so rapidly it is to be expected that our
review \mbox{overlooks} some work, and all the contributions helpfully
supple\-ment our paper. Our remarks focus on points of possible
disagreement or where expansion seems useful.\looseness=-1

\section{Cooley and Sain}

Cooley and Sain bring to the discussion wide experience of statistical
applications in atmospheric science, in addition to innovative
methodological work. We entirely agree with them that the analysis of
annual or seasonal maxima is often unsatisfactory from the statistical
point of view: it fails to make full use of the available data, which
typically comprise numerous simultaneous time series, and by reducing
daily or even hourly data to annual maxima does not allow detailed
modeling of the underlying process. In some cases it is useful to
follow \citet{StephensonTawn2005} and to incorporate information on
the occurrence times of annual maxima; \citet{DavisonGholamrezaee2012}
show that this is quite feasible in the present context, and find some
improvement in precision of estimation from doing so. There is a~close
relationship between models for annual maxima, as\vadjust{\goodbreak} considered in our
paper, and those for peaks over thresholds (\cite
{Smith1989}; \cite{DavisonSmith1990}), and max-stable models of both types share
the deficiencies mentioned at the end of Section~8 of our paper. \citet
{HuserDavison2012} extend the ideas used for annual maxima in the
present paper to a space-time treatment of extreme hourly rainfall data
using the threshold approach. The use of pairwise likelihood poses some
tricky issues in that context, however, because of the multiplicity of
pairs, which can correspond to simultaneous events in different time
series, events at different times in a single series, or at different
times in different series. The application considered by \citet
{HuserDavison2012} involves 10 hourly rainfall time series for 27
summers, around 580,000 observations giving 7 billion possible pairs,
of which a subset of only around 30 million were used! Although heavy
computational burdens arise also in other spatial modeling contexts,
better approaches are clearly needed to deal with larger settings for
spatial extremes, as Cooley and Sain remark. As an aside, the choice of
subsets of observations that contribute to the composite likelihood can
be more subtle than at first appears: \citet{HuserDavison2012} find
that although one might think it best to include only
strongly-dependent pairs, it can be preferable to include some for
which observations are independent or nearly so, in order to get
reasonable estimates of the ranges of extremal phenomena.

We entirely agree that the goals of analysis may differ, and that it
may not be worthwhile to fit a~spatial (or space-time) extremal model
when a map of quantiles is the intended output. However, naive use of a
latent variable model that ignores the correlations between the events
may provide uncertainty measures that are overly precise, as pointed
out in~the discussion contribution by Gabda et al. Thus, building some
form of spatial dependence between events, and not merely between model
parameters, seems wise. A pragmatic way to do this may be the~use of a
Gaussian copula, as in~\citet{SangGelfand2010}.\looseness=-1

\begin{figure*}[b]
\vspace*{-6pt}
\includegraphics{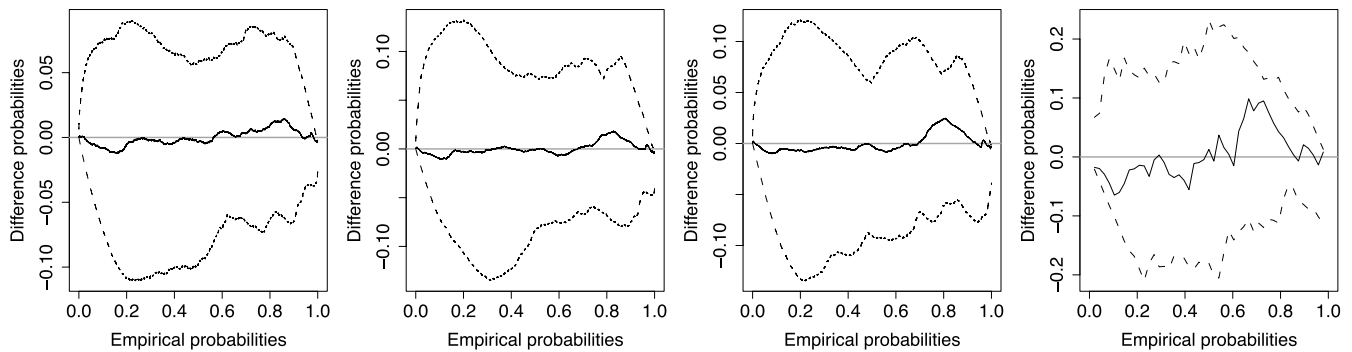}\vspace*{-5pt}
\caption{Rescaled P-P plot for $Z_D$ derived from the annual
maximum rainfall data, with bootstrap pointwise 95\% confidence sets.
From left to right: $k=2$, 3, 4, 36.}
\label{diagfig}
\end{figure*}

Cooley and Sain's final comment concerns a crucial part of extremal
modeling, namely, the incorporation of subject-matter knowledge. While
the generalized extreme-value distribution, max-stable process and the\vadjust{\goodbreak}
like rest on elegant and mathematically compelling theory, the real
world is a messy place to which the relevance of that theory may be
unclear. Very often extremal data show much greater variation than a
simplistic view of the theory might suggest, perhaps due to unsuspected
dependencies or to underlying heterogeneous phenomena. \citet
{Garavagliaetal2010} and \citet{SuvegesDavison2012} suggest two
approaches to modeling in such cases, the first based on a
decomposition into weather types, with a different extremal model
fitted for each, and the second using a more conventional statistical
approach based on a mixture model. If sufficiently full data are
available, the first approach incorporates substantive information more
fully and therefore seems preferable, but the second may be a~useful backstop.\vspace*{-2.5pt}

\section{Gabda, Towe, Wadsworth and Tawn}\vspace*{-2.5pt}

Tawn and his co-workers have made many innovative
contributions to statistics of extremes, and their discussion
contribution does not disappoint. They are quite correct to say that
despite its mathematical attraction, the max-stability of classical
multivariate extreme-value distributions is often inappropriate for
data, and for that reason we welcome their suggested diagnostic, which
is useful beyond the spatial setting. It is related to plots such as
Figure~5 of the paper, which compares observed maxima for selected
groups of stations with corresponding maxima for simulations from a
fitted max-stable model. The main difference is that the proposed
diagnostic compares maxima for data from many more groups of stations,
all of the same sizes, directly with a fitted Gumbel distribution. One
might therefore expect it to have greater power, and to get a feeling
for this we applied it to our data. These have both fewer replicates
and fewer stations than used in the simulations\vadjust{\goodbreak} of Gadba et al., who
generate 1000 replications at 100 stations on a grid, whereas we have
47 replications at 36 irregularly-spaced stations in the data we use
for fitting. The resulting diagnostic plots, with $|D|=2$, 3, 4, 36,
are shown in Figure~\ref{diagfig}. We use all 630 pairs of sites for
$|D|=2$, 620 randomly-chosen triplets and quadruplets for $|D|=3$ and
4, and, of course, just one set when $|D|=36$. Although the power of
the diagnostic will be much lower than in the simulations of Gadba et
al., the figure does not suggest that the max-stability assumption is
unreasonable for our data---indeed, it seems to give a surprisingly good fit.
In other settings we have mixed experience with rainfall data: in a
very detailed but short-term data set from a high Alpine watershed,
max-stability seems appropriate for stations just a few hundred meters
apart, while in a longer-term data set from South Africa,
near-independence seems to apply, though at longer distances.

There is clearly scope for further investigation here, and for the
construction of different diagnostics, for example, developing ideas of
\citet{NaveauGuillouCooleyDiebolt2009} beyond max-stability.\vspace*{-2.5pt}

\section{Segers}\vspace*{-2.5pt}

Segers has made some important theoretical contributions, and his
discussion nicely supplements our rather incomplete treatment of the
copula approach to modeling extremes. While we agree that the
nonparametric methods he describes are valuable for exploratory
analysis and for assessing the quality of fit of parametric models, we
feel that further development is likely to be needed before they can be
routinely used in settings like that discussed in our paper. One reason
for this is the apparent restriction to max-stable models. Although
such models seem to fit our data, they may be unrealistic in other
settings, and spatial models for near-independence (\cite
{WadsworthTawn2012spatial}) are\vadjust{\goodbreak} clearly an important development that
greatly enlarges practical modeling possibilities. More fundamentally,
we can't see how the nice theory for estimating the Pickands dependence
function that Segers describes deals with the spatial element: how does
one make predictions at ungauged stations, or simulate realizations for
entire regions, based on nonparametric fits for the data at gauged
stations? Moreover, and since the available time series in
environmental applications are often much shorter than the periods for
which extrapolation is required, why would it be useful to avoid
modeling the marginal behavior, since in that case extrapolation beyond
the data would not be possible? Finally, we suspect that
a~nonparametric estimate of a 35-dimensional extremal dependence function
based on 47 independent observations, as would be necessary in our
application, is unlikely to be useful---it seems clear that some sort
of strong structural constraints would have to be applied, perhaps
using ideas of empirical likelihood (\cite{EinmahlSegers2009}).\vspace*{-2.5pt}

\section{Shaby and Reich}\vspace*{-2.5pt}

These authors' Bayesian approach to fitting the Smith model is a nice
contribution, and its appearance in print will certainly stimulate work
on fitting more general models---as mentioned above, fitting methods
that can deal with the complexities and size of climatic and
atmospheric data sets are badly needed for applications. We entirely
agree that viewing extreme-value problems, where moments are of
doubtful utility, through Gaussian spectacles, is not typically
helpful, and that scale-invariant quantities such as the Brier score
are more valuable. Our feelings about quantile regression are mixed:
\citet{NorthropJonathan2011} suggest using this approach to specify a
covariate-dependent threshold, but as mentioned in its discussion
(\cite{ChavezDavisonFrossard2011}), it seems to us that in many cases
it will be preferable to base inference on the largest few order
statistics at each site; this implicitly specifies a threshold but
without having to fit a quantile regression model separate from a peaks
over threshold model, and then having to pull together uncertainties
from these two estimators. Using a Bayesian approach for individual
series seems reasonable, though expanding a model using any available
substantive knowledge seems preferable to naive use of a nonparametric
Bayes approach, but the difficulty in specifying
asymptotically-justified joint densities for extremes at different
sites again raises its ugly head, if a model consistent with known
theory for extremes is required.\vadjust{\goodbreak}

\section{Conclusion}

Given the degree of current research activity in the area, it seems
reasonable to hope that some of the problems raised above will be
solved fairly soon. This is devoutly to be wished, since flexible but
mathematically justified inferences for spatial and spatio-temporal
extremes are urgently needed in applications.

We thank the Editor for organizing the discussion.

%

\end{document}